\documentclass[a4paper]{article}

\usepackage[numbers, sort]{natbib}

\usepackage{INTERSPEECH2022}
\usepackage{xspace}
\usepackage{tikz}
\usepackage{algorithm,algorithmicx,algpseudocode}
\usepackage{mathtools}
\usepackage{amsmath, amssymb}
\usetikzlibrary{calc, shapes}
\usetikzlibrary{decorations.text}
\usetikzlibrary{arrows.meta}
\usepackage{hyperref}
\usepackage{url}
\usepackage{multirow}
\usepackage{tabularx}
\usepackage{subcaption}
\usepackage{siunitx,mwe}

\title{NU-Wave 2: A General Neural Audio Upsampling Model\\ for Various Sampling Rates}
\name{Seungu Han$^{1, 2}$, Junhyeok Lee$^1$}

\address{$^1$MINDsLab Inc., Republic of Korea \qquad $^2$Seoul National University,  Republic of Korea}
\email{\{hansw0326, jun3518\}@mindslab.ai}

\makeatletter
\DeclareRobustCommand\onedot{\futurelet\@let@token\@onedot}
\def\@onedot{\ifx\@let@token.\else.\null\fi\xspace}

\def\etc{\emph{etc}\onedot}

\makeatother

\newcommand{\inwt}{\textit{NU-Wave 2}\xspace}
\newcommand{\nwt}{NU-Wave 2\xspace}
\newcommand{\inw}{\textit{NU-Wave}\xspace}
\newcommand{\nw}{NU-Wave\xspace}
\newcommand{\asr}{audio super-resolution\xspace}
\begin{document}

\maketitle

\begin{abstract}
Conventionally, audio super-resolution models fixed the initial and the target sampling rates, which necessitate the model to be trained for each pair of sampling rates. We introduce NU-Wave 2, a diffusion model for neural audio upsampling that enables the generation of 48 kHz audio signals from inputs of various sampling rates with a single model. Based on the architecture of NU-Wave, NU-Wave 2 uses short-time Fourier convolution (STFC) to generate harmonics to resolve the main failure modes of NU-Wave, and incorporates bandwidth spectral feature transform (BSFT) to condition the bandwidths of inputs in the frequency domain. We experimentally demonstrate that NU-Wave 2 produces high-resolution audio regardless of the sampling rate of input while requiring fewer parameters than other models. The official code and the audio samples are available at \url{https://mindslab-ai.github.io/nuwave2}.

\end{abstract}

\setlength{\textfloatsep}{12pt}%

\section{Introduction}
Audio super-resolution, or bandwidth-extension is the task of reconstructing high frequencies to generate high-resolution audio from low-resolution audio.
While previous \asr models \cite{dnnbwe,audiosuperres,tfasr,asr4sr,bwegan,tfilm,spkppgbwe} only focus on reconstructing 16 kHz waveforms, higher sampling rate audio signals are in demand in multimedia such as movies or music.
In order to meet these demands, \inw \cite{nuwave}, an \asr model based on diffusion model \cite{dpm, slicedscorematching, ddpm, diffwave, wavegrad}, successfully generates 48 kHz waveforms and naturally reconstructs high-frequency sibilants and fricatives.
However, \citet{nuwave} mentioned that the model cannot generate harmonics of vowels as well. 
Moreover, \nw has been reported to achieve low performance in generating various frequency bands \cite{tunet, voicefixer}.
Our work focuses on overcoming \nw's limitations.

Prior works \cite{dnnbwe,audiosuperres,tfasr,asr4sr,bwegan,tfilm,spkppgbwe,wsrglow,nuwave} fix the initial and the target sampling rates.
However, in real-world application scenarios, it cannot be guaranteed that input speech has identical bandwidth.
Thus, each \asr model must be trained separately and needs to be adapted for each pair of sampling rates.
To address the above problem, we aim to generate high-resolution audio signals of a fixed sampling rate from inputs of various sampling rates, so that high-quality audio signals can be obtained from low-quality inputs of any sampling rate.
We call it as \textit{general neural audio upsampling} task.

In this paper, we present \inwt, a diffusion model for general neural audio upsampling.
First, to address \nw's problems, we apply \textit{short-time Fourier convolution} (STFC), the modified fast Fourier convolution (FFC) \cite{ffc}, since FFC performs well for capturing periodic signals \cite{lama}.
Furthermore, inspired by the works on feature normalization in image domain \cite{sft, spade}, we propose a simple bandwidth conditioning method operating in the spectral domain, named \textit{bandwidth spectral feature transform} (BSFT), for adapting to various sampling rates.
Our method generates high-quality audio signals from inputs of various sampling rates with fewer parameters than other models.

\section{Diffusion Modeling} \label{section:dpm}
Diffusion models are latent variable models comprised of a diffusion process $\{z_t\}^1_{t=0}$ with time $t \in [0,1]$, which starts from the marginal distribution of data $p(x)$.
The forward process is a diffusion process that samples a sequence of latent $z_t$ where time flows from $t=0$ to $t=1$. Conventionally, the forward process $q(z_t|x)$ is given as the following Gaussian distribution:
\begin{align}\label{eqn:zt}
q(z_t|x) = \mathcal{N}(\alpha_t x, \sigma^2_t I),
\end{align}
where $\alpha_t$ and $\sigma^2_t$ are specified by a noise schedule.
The noise schedule $\lambda_t = \log[\alpha^2_t / \sigma^2_t]$ is defined as the log signal-to-noise ratio (SNR), which is a function that strictly monotonically decreases in $t$.
In this paper, we choose the variance preserving diffusion process \cite{dpm, ddpm}, where $\sigma_t^2 = 1- \alpha_t^2$. Then, $\alpha_t^2$ and $\sigma_t^2$ can be reparametrized by $\lambda_t$ as
$\label{eqn:atst}
    \alpha_t^2 = \text{sigmoid}(\lambda_t), \sigma_t^2 = \text{sigmoid}(-\lambda_t)$.
We use the same noise schedule from \citet{classifierdiff} by setting $\lambda_t = -2 \log \tan{(at + b)}$, where $b = \arctan{(e^{-\lambda_0/2})}$ and $a = \arctan{(e^{-\lambda_1/2})} - b$ with $\lambda_0 = 20$ and $\lambda_1 = - 20$.
With the noise schedule we set, we get $\alpha_t = \cos{(at + b)}$ and $\sigma_t = \sin{(at + b)}$.

By inverting the forward process, which gives the reverse process, we can sample $z_0 \sim p(x)$.
With Eq. \eqref{eqn:zt}, the distributions $q(z_s|z_t, x)$ are also Gaussian by Bayes rule for any $0 \le s < t \le 1$.
For our case, the reverse process can be performed by feeding a denoising model $x_{\theta}(z_t, \lambda_t)$ into $q(z_s | z_t, x)$, starting from $z_1 \sim \mathcal{N}(0, I)$. On the other hand, $x_{\theta}(z_t, \lambda_t)$ can be parametrized in terms of a noise prediction model $\epsilon_\theta(z_t, \lambda_t)$ as $x_{\theta}(z_t, \lambda_t) = (z_t - \sigma_t \epsilon_{\theta}(z_t, \lambda_t))/\alpha_t$. \citet{ddpm} suggests minimizing the following unweighted mean squared error results in better generation quality compared to predicting $x$ directly:
\begin{align}\label{eqn:loss}
     \mathbb{E}_{\epsilon, t}\left[\lVert 
     \epsilon - \epsilon_\theta(z_t, \lambda_t) 
     \rVert_2^2\right],
\end{align}
where $\epsilon \sim \mathcal{N}(0,I)$. 
For the $k$ samples within the minibatch, we sample $u \sim \mathcal{U}([0, 1])$ and then apply $t^i$ as $mod(u+i/k, 1)$, where $i \in {1,2,\ldots, k}$. 
This makes the samples' times to cover $[0,1]$ more evenly compared to sampling them independently.
To reduce variance and to improve training stability, we train the diffusion model using $L^1$ norm instead of $L^2$ norm \cite{wavegrad}.
Algorithm \ref{alg:training} illustrates the training procedure for \nwt. The model receives the latent $z_t$, the log-SNR $\lambda_t$, the low-resolution signal $x_l$ and the bandwidth embedding $e_l$ as inputs. Bandwidth embedding is described in Section \ref{section:arch}.

\begin{figure}[t]
\setlength{\intextsep}{0pt}%
\begin{minipage}[t]{\linewidth}
\begin{algorithm}[H]
  \caption{\small Training.} \label{alg:training}
  \begin{algorithmic}[1]
    \algrenewcommand\algorithmicindent{1.0em}
    \Repeat
      \State $x \sim q(x), u \sim \mathcal{U}([0, 1])$
      \State $t^i = mod(u+i/k,1) \text{ for batch size } k, i \in \{1,2,\ldots, k\}$
      \State $x_l = \text{Upsampling(Downsampling}(\text{Filtering}(x)))$
      \State $b = \arctan{(e^{-\lambda_0/2})}, a = \arctan{(e^{-\lambda_1/2})} - b$
      \State $\lambda_t = -2 \log \tan{(at + b)}$
      \State $\alpha_t = \cos{(at + b)}, \sigma_t = \sin{(at + b)}$
      \State $\epsilon \sim \mathcal{N}(0,I)$
      \State Take gradient descent step on
      \Statex $\quad\; \nabla_\theta\! \lVert \epsilon - \epsilon_\theta(\alpha_t\,x + \sigma_t\,\epsilon, x_{l}, e_l, \lambda_t) \rVert_1$
    \Until{converged}
  \end{algorithmic}
\end{algorithm}
\end{minipage}

\setlength{\intextsep}{10pt}%
\begin{minipage}[t]{\linewidth}

\begin{algorithm}[H]
  \caption{\small Sampling.}%
  \label{alg:sampling}
  \begin{algorithmic}[1]
    \algrenewcommand\algorithmicindent{1.0em}
    \State $z_{t_N} \sim \mathcal{N}(0, I), t_0=0$
    \For{$i=N, N-1, \dotsc, 1$}
      \State $x_{\theta}(z_{t_i}, x_{l}, e_l, \lambda_{t_i}) =  (z_{t_i} - \sigma_{t_i} \epsilon_\theta(z_{t_i}, x_{l}, e_l, \lambda_{t_i}))/\alpha_{t_i}$
      \State $z_{t_{i-1}} = \alpha_{t_{i-1}} x_{\theta}(z_{t_i}, x_{l}, e_l, \lambda_{t_i})$ 
      \Statex \hspace{2.5cm} $+ \sigma_{t_{i-1}} \epsilon_\theta(z_{t_i}, x_{l}, e_l, \lambda_{t_i})$
    \EndFor
    \State \textbf{return} $\hat x = z_0$
  \end{algorithmic}
\end{algorithm}
\end{minipage}
\end{figure}

Since Eq. \eqref{eqn:loss} is denoising score matching, our model can infer the gradient for the log probability density of the latent, which can be expressed as $\epsilon_\theta \approx - \sigma_t \nabla_{z_t} \log(p(z_t))$.
Our model $\epsilon_{\theta}$ then corresponds to a deterministic process satisfying the probabilistic flow ODE \cite{sde}:
\begin{align}
    dz = \left[f(t) z_t - \frac{1}{2} g^2(t) \nabla_{z_t} \log(p(z_t))\right] dt,
\end{align}
where time $t$ flows backwards from $1$ to $0$, and
$f(t) = \frac{d \log \alpha_t}{dt}, \quad g^2(t) = \frac{d\sigma^2_t}{dt} - 2 \frac{d \log \alpha_t}{dt} \sigma^2_t$ \cite{vdm}.

As shown by \citet{progressive}, the DDIM \cite{ddim} sampler can also be understood as an integration rule for the probability flow ODE.
For $s<t$, DDIM uses the following update rule:
\begin{align}\label{eqn:ddim}
    z_s = \alpha_s \frac{z_t - \sigma_t \epsilon_\theta}{\alpha_t} + \sigma_s \epsilon_\theta = \alpha_s x_\theta + \sigma_s \epsilon_\theta.
\end{align}

During inference, we sample with the DDIM update rule. The inference speed depends on the number of iterations $N$, and we set $N$ to 8 for fast sampling, the same as \nw \cite{nuwave}. After trial and error, we set the noise schedule to $\lambda_{t_{1:8}} = [-2.6, -0.8, 2.0, 6.4, 9.8, 12.9, 14.4, 17.2]$. 
We do not report the experimental results with large number of iterations, since the results are similar to those obtained with 8 iterations.
Algorithm \ref{alg:sampling} displays the sampling procedure for \nwt.

\section{Model Architecture}\label{section:arch}
\begin{figure}[t!]
  \centering
  \includegraphics[width=.85\linewidth]{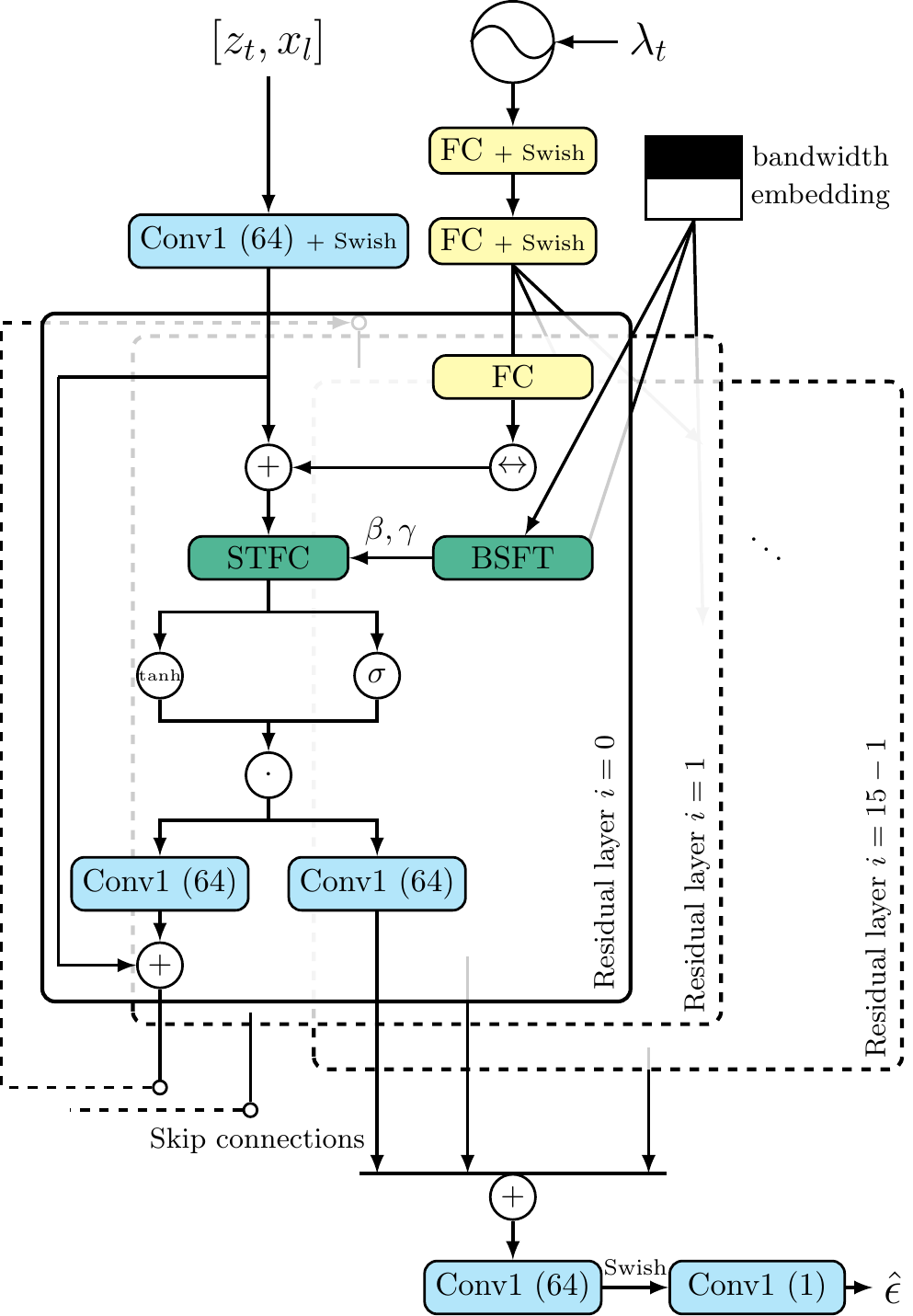}
        \vspace{-0.5\baselineskip}
  \caption{The network architecture of \nwt. Noisy waveform $z_t$, low-resolution speech $x_{l}$, bandwidth embedding of $x_l$, and log-SNR $\lambda_t$ are inputs of model. The model estimates noise $\hat{\epsilon}$ to reconstruct $x$ from $z_t$.}
  \label{fig:model_figure}
\end{figure}

\begin{figure}[t!]
  \centering
  \includegraphics[width=.95\linewidth]{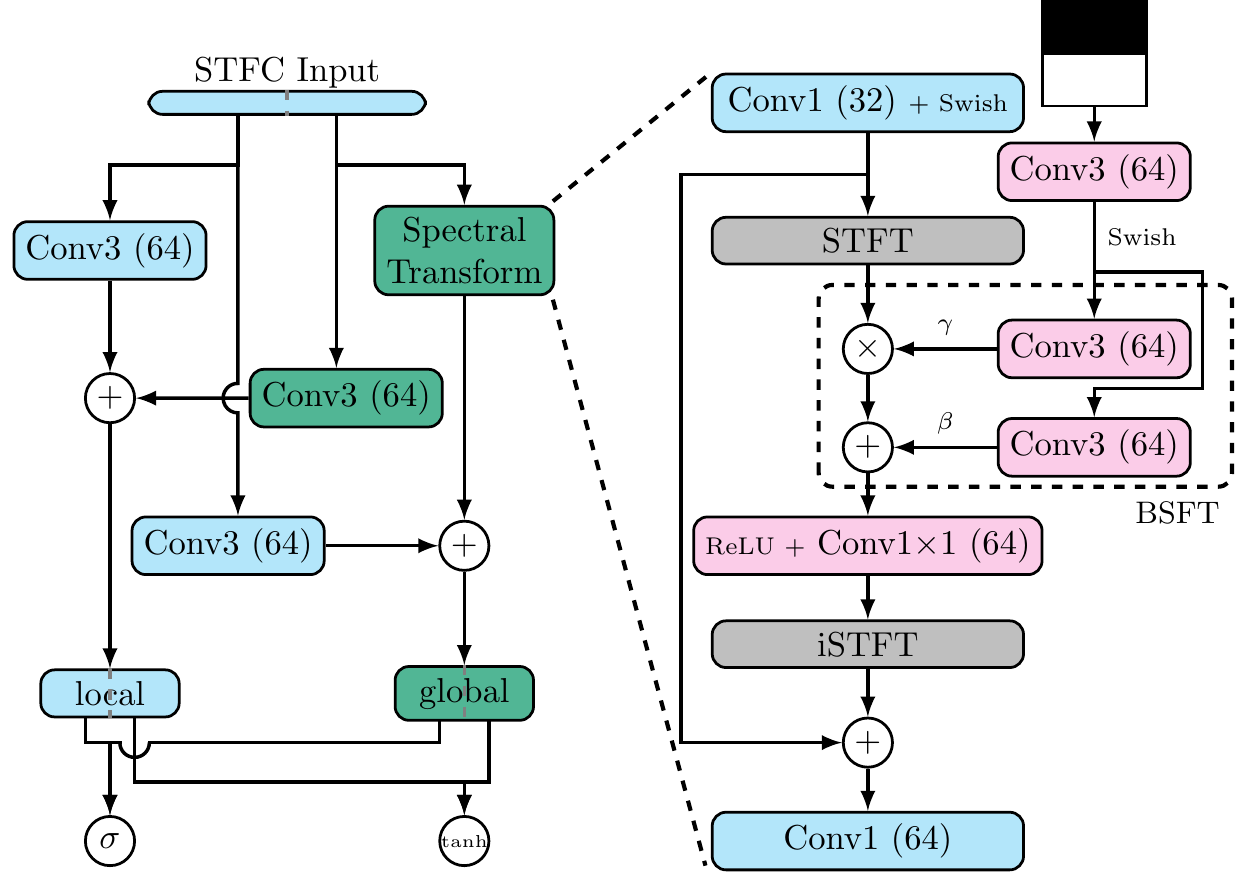}
        \vspace{-0.5\baselineskip}
  \caption{Architecture design of short-time Fourier convolution (STFC) and spectral transform with bandwidth spectral feature transform (BSFT).}
  \label{fig:ffc_bsft}
\end{figure}

Compared to \nw \cite{nuwave}, \nwt concentrates on generating harmonics and adapting to various bandwidths of inputs. \nwt is based on the architecture of \nw with some modifications. \nwt takes the noisy speech $z_t$, the low-resolution signal $x_l$, the log-SNR $\lambda_t$ and the bandwidth embedding $e_l$.
Bandwidth embedding is a one-hot embedding that labels whether each band originates from the low-resolution signal $x_l$ or not. To condition $x_l$, we follow \citet{sr3}, and just concatenate $x_l$ with $z_t$ along the channel dimension. Similar to \nw, our model is composed of $N=15$ residual layers with 64 channels.
 In each residual layer, to generate harmonics and resolve the main failure mode of \nw, we use \textit{short-time Fourier convolution} (STFC), the modified fast Fourier convolution (FFC) \cite{ffc}, which consists of a local and a global branches.
When applying to gated-tanh nonlinearities, the outputs of the local and the global branches are split for each along the channel dimension and concatenate in pairs based on the channel dimension.
Furthermore, we propose \textit{bandwidth spectral feature transform} (BSFT) that conditions the bandwidth of input audio by transforming the features in the spectral domain. 
Figure \ref{fig:model_figure} shows the overall architecture of \nwt and Figure \ref{fig:ffc_bsft} illustrates STFC and BSFT.

\subsection{Short-time Fourier convolution (STFC)}
Short-time Fourier convolution (STFC) consists of two parts, a local branch that conducts conventional convolution and a global branch with a spectral transform that performs convolution in the frequency domain.
We split STFC input exactly half along the channel dimension.
Although original FFC has been proposed only considering image data, we modify FFC to fit audio data.
In FFC, the global branch uses real fast Fourier transform (FFT) \cite{fft} to transform original features into the frequency domain.
Since the frequency resolution of FFT is affected by input shape, FFT cannot be applied to data with varying lengths such as audio.
Instead of FFT, we use short-time Fourier transform (STFT) in STFC to convert inputs to the spectral features.
We concatenate the real part and the imaginary part of the STFT output in the channel dimension.
After that, we apply a $1 \times 1$ convolution on the spectral features, split them into the real and the imaginary part, and convert them back to the temporal domain using inverse short-time Fourier transform (iSTFT). Finally, the outputs of the local and the global branches are fused together.

Each STFC provides a high receptive field as window size which is vital for high-resolution data.
We set STFT size to 1024, hop size to 256, and Hanning window of size 1024. 
Since FFCs are capable of capturing periodic signals \cite{lama}, we use STFCs in each residual layer to generate periodic signals and reconstruct harmonics.

\subsection{Bandwidth spectral feature transform (BSFT)}
We present bandwidth spectral feature transform (BSFT) layer for conditioning various bandwidths of input low-resolution audio simply for adapting to various sampling rates.
Previous works \cite{sft, spade} on feature normalization in image domain inspired our work. While they deal with preserving semantic information in the spatial domain, we work on preserving the input low frequencies in the spectral domain. We can give the input bandwidth as a condition after the features have been transformed into the spectral domain in STFCs.
By giving the bandwidth condition directly in the spectral domain, the model can determine where to conserve and where to generate according to the frequency band. 
The transformation parameters $\gamma$ and $\beta$ are obtained from the bandwidth one-hot embedding which is a 2-D tensor size of $2 \times F$, where $F$ is the frequency dimension of the STFT output.
The transformation is then performed by scaling and shifting features of a specific layer: $\gamma \odot h + \beta$, where $h$ denotes the spectral features. $\gamma$ and $\beta$ are obtained for each layer using two convolutions. As the input bandwidth boundary can be unclear depending on the low pass filter, we used convolutions with kernel size 3.

The proposed BSFT layer allows conditioning bandwidth in an uncomplicated way with only bandwidth information of input audio.
We show that the model preserves low frequencies that are almost identical to ground truth.

\section{Experiments}
\subsection{Training}\label{subsection:train}
We train the model on VCTK dataset \cite{vctk}, which includes 44 hours of recorded speech by 108 speakers. For the speakers, we follow the training and the test setup from \nw \cite{nuwave}.
The target sampling rate is 48 kHz and the input sampling rate is uniformly sampled from $[6000, 48000]$ for training.
We generate training data by applying the Chebyshev Type I low-pass filter with random ripples and orders for generalization which achieves robustness to different low-pass filters \cite{tunet}.
After the filtering, the audio signals are downsampled to the input sampling rate, then upsampled back to 48 kHz for setting the lengths regardless of the input sampling rates.
For testing, we test on 8 kHz, 12 kHz, 16 kHz, and 24 kHz input data using a single 8th order Chebyshev Type I low-pass filter with 0.05 dB of ripple.

Our model is trained using the Adam \cite{adam} optimizer with the learning rate $2\times10^{-4}$. For training, we use the 0.682 seconds patches (32768 samples) from the signals. We set the batch size to 24 for all models.

\subsection{Evaluation}
We evaluate our results quantitatively by measuring the signal-to-noise ratio (SNR) and log-spectral distance (LSD).
Several works say that high-frequency generation cannot be measured with SNR, therefore SNR is not suitable for upsampling task \cite{audiosuperres,bwegan}. On the other hand, LSD could be used to measure high-frequency generation as spectral distance.
Additionally, we measure high-frequency LSD (LSD-HF) and low-frequency LSD (LSD-LF), which are the LSD values calculated for the reconstructed high-frequency range and input low-frequency range respectively.
While the conservation of low frequencies cannot be verified from LSD independently, 
LSD-LF is able to measure the preservation of low frequencies.

To determine whether the generated results can be distinguished from the ground truth, we perform ABX test for qualitative measurement.
ABX test is hosted on Amazon Mechanical Turk system and we examined almost 1800 cases.

\subsection{Baselines}
We compare our method with WSRGlow \cite{wsrglow} and \nw, which are aiming for high-quality 48 kHz audio.
To train baseline models for general neural audio upsampling task, we use the same VCTK dataset and the same method to generate low-resolution audio signals as in Section \ref{subsection:train}.
While \nw can be trained without modification, original WSRGlow requires some adjustment due to input length issues caused by downsampled input audio not passing through an upsampling process.
As the lengths of inputs are unified to the original signals for general neural audio upsampling task, we add a convolution layer with kernel size 3 and stride 2 after conditional encoder of $2\times$ upsampling WSRGlow model.
While adjusted WSRGlow's number of parameters is 270M, the number of parameters for \nw and \nwt are significantly small, 3.0M and 1.7M, respectively.
Compared to other baselines, \nwt has the fewest parameters, even fewer than \nw.
There is a possibility that hyperparameters of baseline models are not fully optimized for our experimental setting.

\begin{figure}[th!]
\setlength{\textfloatsep}{0pt}%

\newcommand{\incaptionimg}[3]{
  \begin{tikzpicture}[every node/.style={inner sep=0,outer sep=0}]
    \draw node[name=micrograph] {\includegraphics[width=\textwidth]{#1}}; 
    \draw  (micrograph.north west)  node[anchor=north west,yshift=-1.75cm,xshift=0.40cm,#3]{\textbf{\small{(#2)}}}; 
  \end{tikzpicture}
}
\newcommand{\incaptionlineimg}[4]{  
  \begin{tikzpicture}[every node/.style={inner sep=0,outer sep=0}]
    \draw node[name=micrograph] {\includegraphics[width=\textwidth]{#1}}; 
    \draw[line width =0.5pt, red] (micrograph.south west)++(0.093*\textwidth,#4\textwidth)--++(0.907\textwidth,0); 
    \draw  (micrograph.north west)  node[anchor=north west,yshift=-1.75cm,xshift=0.40cm,#3]{\textbf{\small{(#2)}}}; 
  \end{tikzpicture}
}
\captionsetup[subfigure]{labelformat=empty}
    \setlength{\textfloatsep}{0pt}%
    \setlength{\intextsep}{0pt}
    \centering
    \vspace{-0.8\baselineskip}
     \begin{subfigure}{0.49\linewidth}
         \subcaption{\quad Reference Signal}
         \vspace*{-0.75em}
         \incaptionimg{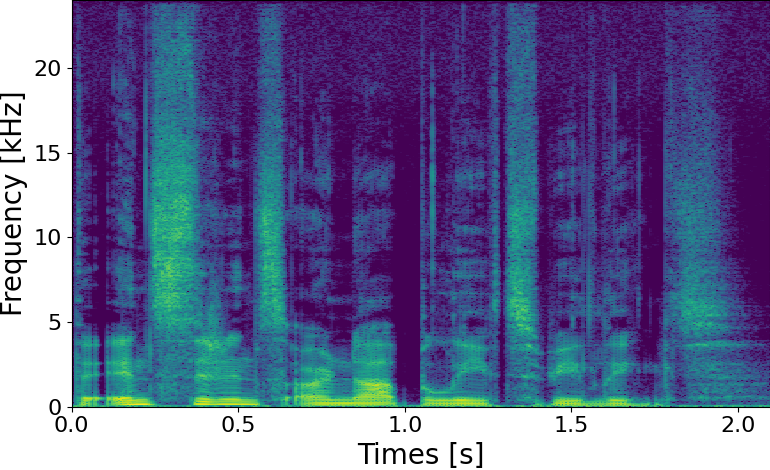}{a}{white}
         \phantomsubcaption\ignorespaces\label{spec_a}
     \end{subfigure}%
        \hspace*{-0.1em}
    \begin{subfigure}{0.49\linewidth}
         \subcaption{\quad NU-Wave}
         \vspace*{-0.75em}
         \incaptionlineimg{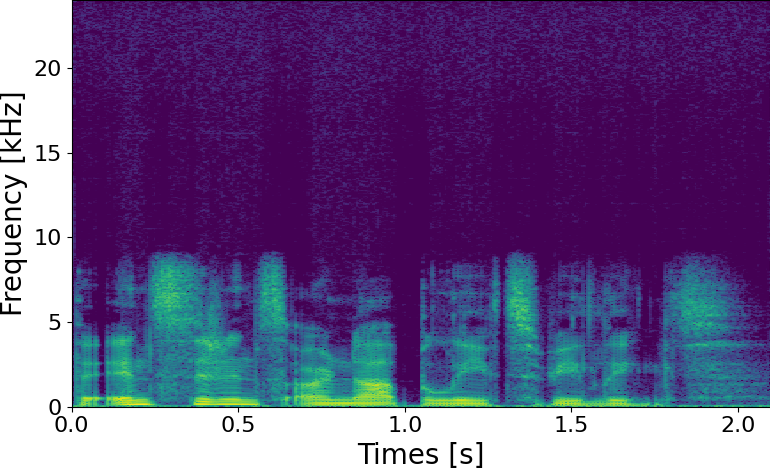}{b}{white}{0.260}
         \phantomsubcaption\ignorespaces\label{spec_b}
     \end{subfigure}%
     \vspace{-1.0\baselineskip}
    \begin{subfigure}{0.49\linewidth}
         \subcaption{\quad WSRGlow}
         \vspace*{-0.75em}
         \incaptionlineimg{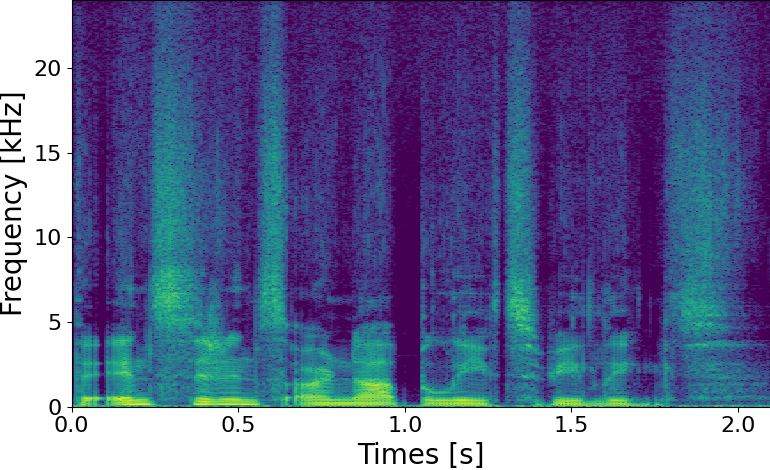}{c}{white}{0.260}
         \phantomsubcaption\ignorespaces\label{spec_c}
     \end{subfigure}%
          \hspace*{-0.1em}
     \begin{subfigure}{0.49\linewidth}
         \subcaption{\quad NU-Wave 2 (ours)}
         \vspace*{-0.75em}
         \incaptionlineimg{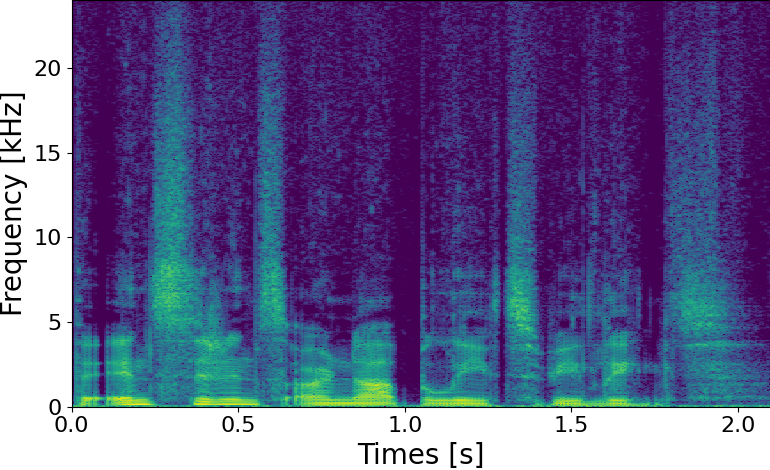}{d}{white}{0.260}
         \phantomsubcaption\ignorespaces\label{spec_d}
     \end{subfigure}%
    \vspace{-1.8\baselineskip}
    \caption{Spectrograms of reference and upsampled speeches (p360\_059) for 16 kHz input. Red lines indicate the Nyquist frequencies of downsampled signals.
    }
    \label{fig:spec}
\end{figure}

\begin{figure}[th!]
\setlength{\textfloatsep}{0pt}%

\newcommand{\incaptionimg}[3]{
  \begin{tikzpicture}[every node/.style={inner sep=0,outer sep=0}]
    \draw node[name=micrograph] {\includegraphics[width=\textwidth]{#1}}; 
    \draw  (micrograph.north west)  node[anchor=north west,yshift=-1.75cm,xshift=0.35cm,#3]{\textbf{\small{(#2)}}}; 
  \end{tikzpicture}
}
\newcommand{\incaptionlineimg}[4]{  
  \begin{tikzpicture}[every node/.style={inner sep=0,outer sep=0}]
    \draw node[name=micrograph] {\includegraphics[width=\textwidth]{#1}}; 
    \draw[line width =0.5pt, red] (micrograph.south west)++(0.086*\textwidth,#4\textwidth)--++(0.914\textwidth,0); 
    \draw  (micrograph.north west)  node[anchor=north west,yshift=-1.75cm,xshift=0.35cm,#3]{\textbf{\small{(#2)}}}; 
  \end{tikzpicture}
}
\captionsetup[subfigure]{labelformat=empty}
    \setlength{\textfloatsep}{0pt}%
    \setlength{\intextsep}{0pt}
    \centering
     \begin{subfigure}{0.49\linewidth}
         \subcaption{\quad Reference Signal}
         \vspace*{-0.80em}
         \incaptionimg{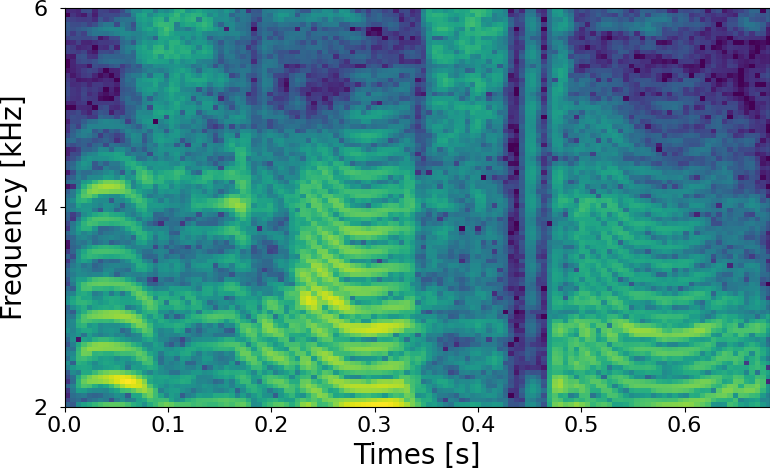}{a}{white}
         \phantomsubcaption\ignorespaces\label{spec_a_har}
     \end{subfigure}%
        \hspace*{-0.1em}
    \begin{subfigure}{0.49\linewidth}
         \subcaption{\quad NU-Wave}
         \vspace*{-0.80em}
         \incaptionlineimg{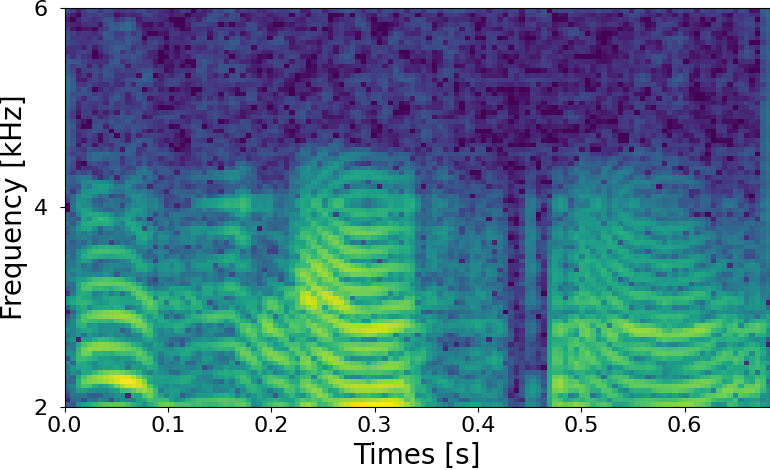}{b}{white}{0.340}
         \phantomsubcaption\ignorespaces\label{spec_b_har}
     \end{subfigure}%
     \vspace{-1.0\baselineskip}
    \begin{subfigure}{0.49\linewidth}
         \subcaption{\quad WSRGlow}
         \vspace*{-0.80em}
         \incaptionlineimg{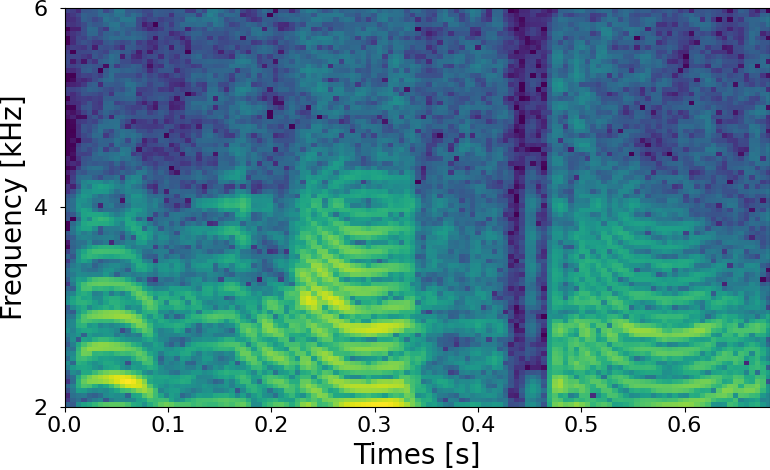}{c}{white}{0.340}
         \phantomsubcaption\ignorespaces\label{spec_c_har}
     \end{subfigure}%
          \hspace*{-0.1em}
     \begin{subfigure}{0.49\linewidth}
         \subcaption{\quad NU-Wave 2 (ours)}
         \vspace*{-0.80em}
         \incaptionlineimg{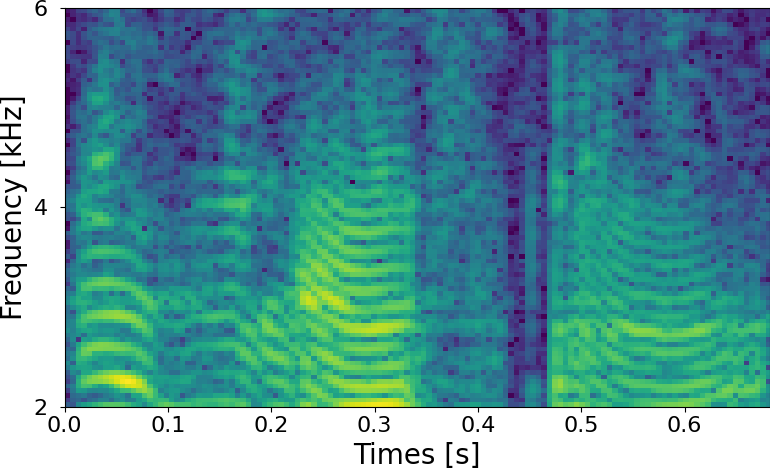}{d}{white}{0.340}
         \phantomsubcaption\ignorespaces\label{spec_d_har}
     \end{subfigure}%
    \vspace{-1.8\baselineskip}
    \caption{Spectrograms zoomed in on 2 kHz to 6 kHz frequency of reference and upsampled speeches (p362\_119) for 8 kHz input. The Nyquist frequencies of inputs are denoted by red lines.
    }
    \label{fig:harmonic}
\end{figure}

\section{Results}
As shown in Figure \ref{fig:spec}, our model can reconstruct valid high frequencies, while \nw cannot. On the other hand, WSRGlow does produce high frequencies but its output is smoothed and noisy.
In Figure \ref{fig:harmonic}, we observe that \nwt is capable of generating harmonics while baselines cannot reconstruct harmonics. They only leave spectral alias of the input's harmonics.

The quantitative evaluation of our model and the baselines is shown in Table \ref{tab:quant}.
For most cases, \nwt outperforms \nw. Although, the LSD-LF and the SNR are comparable, \nwt achieves 24.5\% of input's LSD and LSD-HF on average, while \nw only achieves 40.0\%.
On the other hand, WSRGlow obtains 22.4\% of input's LSD and LSD-HF on average.
However, \nwt is superior to other baselines for the LSD-LF and the SNR.
In terms of the LSD-LF, WSRGlow exceeds 63.7\% of input on average, while \nwt exceeds 40.5\%.
For the inputs of 8 kHz, 12 kHz, 16 kHz, and 24 kHz, our model improves SNR value by 0.2 dB, 0.4 dB, 1.4 dB, and 1.8 dB from WSRGlow, respectively.

\begin{table}[th]
  \caption{Results of quantitative metrics. Sampling rate (SR) is indicated as 8 kHz, 12 kHz, 16 kHz, and 24 kHz.
  Our model achieves close LSD and LSD-HF results compared with WSRGlow and outperforms other baselines for LSD-LF and SNR.
  }
    \vspace{-0.75\baselineskip}

  \label{tab:quant}
  \centering
  \resizebox{0.9\linewidth}{!}{%
  \begin{tabular}{ll cccc}
  \toprule
  \textbf{SR} & \textbf{metrics} & Input & WSRGlow & \nw & \nwt \\
    \midrule
    \multirow{2}{*}{}
    & \#\textbf{params} $\downarrow$& & $270\mathrm{M}$ & $3.0\mathrm{M}$ & $\mathbf{1.7\mathrm{M}}$
    \\
    \midrule
    \multirow{2}{*}{\textbf{8 kHz}}
    & LSD $ \downarrow $ & $4.42$ & $\mathbf{1.05}$ & $1.84$ & $\mathbf{1.14}$
    \\
    & LSD-HF $ \downarrow $ & $4.83$ & $\mathbf{1.14}$ & $2.01$ & $\mathbf{1.24}$
    \\
    & LSD-LF $ \downarrow $ & $0.190$ & $0.291$ & $0.243$ & $\mathbf{0.219}$
    \\
    & SNR $ \uparrow $ & $19.0$ & $18.6$ & $18.7$ & $\mathbf{18.8}$
    \\
    \midrule
    \multirow{2}{*}{\textbf{12 kHz}}
    & LSD $ \downarrow $ & $4.09$ & $\mathbf{0.936}$ & $1.66$ & $\mathbf{1.01}$
    \\
    & LSD-HF $ \downarrow $ & $4.72$ & $\mathbf{1.06}$ & $1.91$ & $\mathbf{1.15}$
    \\
    & LSD-LF $ \downarrow $ & $0.206$ & $0.332$ & $0.285$ & $\mathbf{0.275}$
    \\
    & SNR $ \uparrow $ & $22.1$ & $21.2$ & $21.4$ & $\mathbf{21.6}$
    \\
    \midrule
    \multirow{2}{*}{\textbf{16 kHz}}
    & LSD $ \downarrow $ & $3.79$ & $\mathbf{0.842}$ & $1.51$ & $\mathbf{0.925}$
    \\
    & LSD-HF $ \downarrow $ & $4.64$ & $\mathbf{0.995}$ & $1.83$ & $\mathbf{1.10}$
    \\
    & LSD-LF $ \downarrow $ & $0.200$ & $0.353$ & $\mathbf{0.301}$ & $\mathbf{0.305}$
    \\
    & SNR $ \uparrow $ & $24.7$ & $22.6$ & $23.5$ & $\mathbf{24.0}$
    \\
    \midrule
    \multirow{2}{*}{\textbf{24 kHz}}
    & LSD $ \downarrow $ & $3.14$ & $\mathbf{0.701}$ & $1.22$ & $\mathbf{0.774}$
    \\
    & LSD-HF $ \downarrow $ & $4.44$ & $\mathbf{0.924}$ & $1.67$ & $\mathbf{1.02}$
    \\
    & LSD-LF $ \downarrow $ & $0.203$ & $0.333$ & $\mathbf{0.325}$ & $\mathbf{0.326}$
    \\
    & SNR $ \uparrow $ & $29.4$ & $26.6$ & $26.9$ & $\mathbf{28.4}$
    \\
  \bottomrule
  \end{tabular}}
\end{table}

\begin{table}[th]
  \caption{The accuracy and the confidence interval of the ABX test. The accuracy of \nwt is close to 50\% meaning that its outputs are indistinguishable from the reference signals.
  }
    \vspace{-0.75\baselineskip}

  \label{tab:abx}
  \centering
  \resizebox{0.9\linewidth}{!}{%
  \begin{tabular}{l cccc}
  \toprule
  \textbf{SR} & WSRGlow & \nw & \nwt \\
    \midrule
    {\textbf{8 kHz}}
    & $55.0\pm2.35$\% & $58.3\pm2.37$\% & $55.3\pm2.37$\%
    \\
    {\textbf{12 kHz}}
    & $51.3\pm2.37$\% & $56.4\pm2.35$\% & $53.1\pm2.37$\%
    \\
    {\textbf{16 kHz}}
    & $49.3\pm2.39$\% & $52.6\pm2.39$\% & $52.1\pm2.36$\%
    \\
    {\textbf{24 kHz}}
    & $49.0\pm2.36$\% & $50.7\pm2.36$\% & $52.1\pm2.36$\%
    \\
  \bottomrule
  \end{tabular}}
\end{table}

Table \ref{tab:abx} shows the results of the ABX test. While WSRGlow achieves the lowest accuracy, the difference between \nwt was not significant. For each sampling rate, the accuracy of our model is 55.3\%, 53.2\%, 52.1\%, and 52.1\% which are close to 50\%. Therefore, we can say that the output signals of \nwt are nearly indistinguishable from the reference signals.

\section{Discussion}
We proposed a diffusion model for general neural audio upsampling.
Although \nw fails in general neural audio upsampling task, \nwt, an improved model based on \nw, succeeds and outperforms the baseline with fewer parameters.
While \nw was limited in its ability to generate harmonics, \nwt overcomes that limitation by using STFCs and can generate for all frequency bands.
Also, STFCs provide high perceptual quality and parameter efficiency for both by allowing a high receptive field even in the early layers of the network.
While speech bandwidth varies in real-world applications, \nwt can generate high-quality audio regardless of the sampling rate of input speech by using the proposed bandwidth conditioning method, BSFT.

On the other hand, WSRGlow performs better than the other models in LSD and LSD-HF, but worse in LSD-LF and SNR. It means that WSRGlow performs well at generating high frequencies, but is poor at preserving low frequencies, which is an essential precondition for \asr task. 
Also, we observe that WSRGlow is unable to generate harmonics while \nwt can naturally reconstruct harmonics in high-frequency range.
Moreover, by using just 0.63\% of WSRGlow's parameters, \nwt achieves close LSD and LSD-HF results compared with WSRGlow's values and produces the best SNR and LSD-LF. This indicates that \nwt can generate high frequencies and preserves input low frequencies while requiring far fewer parameters than WSRGlow.
Based on the results of the ABX test, our samples are almost indistinguishable from reference signals. However as mentioned in \nw, there was no significant difference between the models because the downsampled signals already contain sufficient information, which has the dominant energy in the audio frequency range.

In this paper, we evaluate our model for specific sampling rates which are commonly used, but \nwt has the ability to upsample audio signals of any sampling rates such as 11.025 kHz, 22.05 kHz, 32 kHz, 44.1 kHz, \etc.
Therefore, our model is also able to generate high-quality audio from other datasets, such as LJSpeech \cite{lj} which is a dataset with 22.05 kHz. 
In further studies, we could aim to restore additional distortions with a single diffusion model.

\section{Acknowledgements}
The authors would like to thank Minho Kim, Sang Hoon Woo, teammates of MINDsLab, and Jinwoo Kim from KAIST for valuable discussions.

\bibliographystyle{IEEEtranN}

\bibliography{main}

\end{document}